\documentclass[preprint]{aastex}

\slugcomment{Accepted by APJL}

\shorttitle{MAGNETIC FLUX ROPE FORMATION}

\shortauthors{Song et al.}

\begin{document}
\title{DIRECT OBSERVATIONS OF MAGNETIC FLUX ROPE FORMATION DURING A SOLAR CORONAL MASS EJECTION}
\author{H. Q. SONG\altaffilmark{1,2}, J. ZHANG\altaffilmark{2}, Y. CHEN\altaffilmark{1}, AND X. CHENG\altaffilmark{3}}

\affil{1 Shandong Provincial Key Laboratory of Optical Astronomy
and Solar-Terrestrial Environment and Institute of Space Sciences,
Shandong University, Weihai, Shandong 264209, China}
\email{hqsong@sdu.edu.cn}

\affil{2 School of Physics, Astronomy and Computational Sciences,
George Mason University, Fairfax, Virginia 22030, USA}


\affil{3 School of Astronomy and Space Science, Nanjing
University, Nanjing, Jiangsu 210093, China}

\begin{abstract}
Coronal mass ejections (CMEs) are the most spectacular eruptive
phenomena in the solar atmosphere. It is generally accepted that
CMEs are results of eruptions of magnetic flux ropes (MFRs).
However, a heated debate is on whether MFRs pre-exist before the
eruptions or they are formed during the eruptions. Several coronal
signatures, \textit{e.g.}, filaments, coronal cavities, sigmoid
structures and hot channels (or hot blobs), are proposed as MFRs
and observed before the eruption, which support the pre-existing
MFR scenario. There is almost no reported observation about MFR
formation during the eruption. In this letter, we present an
intriguing observation of a solar eruptive event occurred on 2013
November 21 with the Atmospheric Imaging Assembly on board the
\textit{Solar Dynamic Observatory}, which shows a detailed
formation process of the MFR during the eruption. The process
started with the expansion of a low-lying coronal arcade, possibly
caused by the flare magnetic reconnection underneath. The
newly-formed ascending loops from below further pushed the arcade
upward, stretching the surrounding magnetic field. The arcade and
stretched magnetic field lines then curved-in just below the
arcade vertex, forming an X-point. The field lines near the
X-point continued to approach each other and a second magnetic
reconnection was induced. It is this high-lying magnetic
reconnection that led to the formation and eruption of a hot blob
($\sim$ 10 MK), presumably a MFR, producing a CME. We suggest that
two spatially-separated magnetic reconnections occurred in this
event, responsible for producing the flare and the hot blob (CME),
respectively.

\end{abstract}


\section{INTRODUCTION}
Magnetic flux rope (MFR), a volumetric current channel with
helical magnetic field winding around its center axial field
lines, plays a key role in solar eruptions manifested as coronal
mass ejections (CMEs) and/or flares. The first direct
observational evidence of the presence of an outstanding MFR in
the sun-Earth system is from the near-Earth in-situ solar wind
observation of the so called magnetic cloud (Burlaga et al. 1981).
According to a statistical study (Vourlidas et al. 2013), at least
$\sim$ 40~\% coronagraphic observations of CMEs show an apparent
flux-rope geometry in the outer corona. However, the direct
detection of MFRs in the lower corona prior to and during the
eruption has been elusive. Therefore, there has been a heated
debate over whether the MFR exists before the onset of CME
eruption or it is formed during the CME eruption (e.g., Forbes
2000; Klimchuk 2001). Many CME models require a pre-existing MFR
and have suggested a variety of initiation processes, including
direct flux injection (e.g,. Chen 1996),  the catastrophic loss of
equilibrium (Forbes \& Isenberg. 1991; Isenberg et al. 1993; Hu et
al. 2003; Chen et al. 2006, 2007), kink (T\"or\"ok et al. 2004) or
torus (Kliem \& T\"or\"ok 2006; Olmedo \& Zhang 2010) instability.
On the other hand, some theoretical models suggest that an MFR be
formed during the CME eruption process through the magnetic
reconnection of coronal arcade (e.g., Miki\'{c} \& Linker 1994;
Antiochos et al. 1999).

Several lines of observations in the lower corona indicate the
presence of MFRs, e.g., filaments/prominences (Rust \& Kumar
1994), coronal cavities (Wang \& Stenborg 2010), sigmoids (Titov
\& D\'{e}moulin 1999; McKenzie \& Canfield 2008), hot channels
(Zhang et al. 2012; Cheng et al. 2012, 2013b; Li \& Zhang 2013)
and/or hot blobs (Cheng et al. 2011; Su et al. 2012; Patsourakos
et al. 2013; Song et al. 2014). The term``blob'' originates from a
2-dimension description of plasmoids either in numerical
simulations or observations. However, it could also describe the
2-D appearance projected onto the plane of the sky of an
intrinsically 3-D MFR structure when observed along the central
axis. All of the features mentioned above can be observed before
the CME onset, thus supporting the pre-existing-MFR scenario. To
the best of our knowledge, direct observations of the MFR
formation from a non-MFR structure have been rare. Cheng et al.
(2011) observed an MFR (hot blob) growth process during the
impulsive phase of a solar eruption. The hot blob was observed
rising from the source active region, stretching the background
magnetic field lines upward. The stretched field lines then
approached toward the center below the blob resulting in magnetic
reconnection. Above the reconnection site, poloidal magnetic
fluxes were injected into the erupting blob, leading to its
growth.

Different from Cheng et al. (2011), we report in this letter a
formation process of a hot blob from an apparently non-MFR
magnetic arcade, instead of the growth from a pre-existing
progenitor blob. The observation was made by the AIA on board
\textit{Solar Dynamic Observatory (SDO)}. The differential
emission measure (DEM) analysis shows its temperature around 10
MK. The eruption of the hot blob led to a CME, which was recorded
by the Large Angle Spectroscopic Coronagraph (LASCO) (Brueckner et
al. 1995) on board \textit{Solar Heliospheric Observatory (SOHO)}
and the Sun Earth Connection Coronal and Heliospheric
Investigation (SECCHI) (Howard et al. 2008) on board \textit{Solar
Terrestrial Relations Observatory (STEREO)}. Further, the eruption
was driven  by  magnetic reconnections  occurred in two spatially
different places, which are found to play different roles: causing
the arcade expansion and producing the blob, respectively. In
Section 2, the instruments are described. We present the
observations and make discussions in Section 3, which are followed
by a summary in Section 4.

\section{INSTRUMENTS}

The AIA on board \textit{SDO} can image the solar atmosphere in 10
narrow UV and EUV passbands with a cadence of 12 seconds, spatial
resolution of 1.2 arcseconds and FOV of 1.3 R$_\odot$. The
temperature response functions of these passbands cover a
temperature range from 0.6 to 20 MK (O'Dwyer et al. 2010; Del
Zanna et al. 2011; Lemen et al. 2012). During an eruption, the
131~\AA\ and 94~\AA\ passbands are sensitive to the hot plasma
from flare regions and erupting flux ropes, while other passbands
are better for viewing dimming regions and cooler leading fronts
(LFs) of eruptions (e.g., Zhang et al. 2012; Cheng et al. 2011;
Song et al. 2014). AIA's high cadence and broad temperature
coverage provide us the opportunity of observing the MFR formation
process during an eruption, and make it possible for constructing
DEM models of corona plasma. Recently, a DEM-based temperature
analysis method has been used successfully to analyze hot blob
temperatures (Cheng et al. 2012, 2014; Song et al. 2014). In this
letter, we continue to use the same method to study the heating
process of the arcades and hot blob. In addition, the SECCHI on
board \textit{STEREO} and LASCO on board \textit{SOHO} provide CME
images in the outer corona from different perspectives. The
\textit{Reuven Ramaty High Energy Solar Spectroscopic Imager
(RHESSI}; Lin et al. 2002) provides the hard X-ray (HXR) spectrum
and imaging of flares.

\section{OBSERVATIONS AND DISCUSSIONS}

\subsection{The Flare -- CME Event}

On 2013 November 21, an M1.2 class soft X-ray (SXR) flare was
recorded by \textit{Geostationary Operational Environmental
Satellite (GOES)}, which started at 10:52 UT and peaked at 11:11
UT. At this time, the separation angle of \textit{SDO} and
\textit{STEREO-A} is $\sim$149$^{\circ}$. The flare location was
at $\sim$S14W89 (NOAA 11893) from the perspective of \textit{SDO}
as shown in AIA 131 \AA \ image of Figure 1(a). The subregions
depicted with dashed and solid squares show the FOV of Figure 1(c)
and Figures 3(b)-(i), respectively. A hot blob was observed and
depicted with the red dotted line. In the STEREO-A EUVI 195 \AA \
image (Figure 1(b)), the flare region was shown on the disk as a
solid square (also the FOV of Figures 2(c)-(d)). During the
process of the hot blob formation and eruption, a cool LF, best
seen in AIA 171 \AA \ in Figure 1(c) (indicated by the blue dotted
line), was formed, likely through the compression of the plasma
surrounding the hot blob. Note that the LF only formed at a later
time in the fast acceleration phase, similar to the formation
sequence of the hot channel and its associated LF reported in
Zhang et al. (2012). The hot blob is completely absent in the cool
171 \AA\ image, as demonstrated in Figure 1(c) where the red
dotted circle indicates the same location of the hot blob shown in
131 \AA \ image in Figure 1(a). LASCO observed the corresponding
CME from 2.2 to 24 R$_\odot$ with a linear speed of 668 km
s$^{-1}$; the CME had a three-part structure: a bright loop-like
LF (blue dotted), a dark cavity, and an embedded bright core (red
dotted), as shown in Figure 1(d). Through inspecting the AIA and
LASCO movies carefully, we conclude that the bright core and LF in
LASCO correspond to the hot blob in 131 \AA \ and LF in 171 \AA,
respectively.

Using magnetic field observations from Helioseismic and Magnetic
Imager (HMI) on board \textit{SDO}, we find that the magnetic
configuration of NOAA 11983 experienced an obvious emergence
process for several days before the eruption (see supplementary
HMI movie online). On November 12, it is classified as Hale Class
$\beta$ (Figure 2(a)), and on November 18, classified as
$\beta$$\delta$ (Figure 2(b)). Figures 2(c) and (d) present the
EUVI observations before and after the eruption, respectively
(also see EUVI movie online). The location of the bright
post-flare arcade and its two footpoints (depicted with red and
blue dots in Figure 2(d)) indicate where the eruption took place.
As shown in Figure 2(b) where the two footpoints are mapped onto
the magnetogram images, the flare and the eruption occurred along
the polarity inversion line (PIL). The evolution of the
magnetogram indicates that there existed a slow shearing motion
along the PIL. Such motion can lead to the arcade expansion upward
as discussed later.

\subsection{Formation of The Hot Blob}

The most interesting aspect of this event is that the hot blob is
formed from a rising loop arcade at the early stage of the
eruption. The sequence of snapshots of the formation process are
presented in Figure 3. Figure 3(a) shows the GOES 1-8 \AA \ SXR
flux, with eight vertical lines indicating the times of the images
in the panels below. The AIA 131 \AA \ images are processed with
an improved radial filter technique (Ma et al. 2011).

At 10:50:08 UT (panel b), the lower coronal loops were observed
clearly with AIA 131 \AA \ , which likely corresponded to the
arcade in Figure 2(d), but the higher arcade (denoted with red
dotted line) overlying the loops were not obvious. Immediately
after the flare onset, the arcade became visible (panel c,
10:52:08 UT). The arcade should have existed before the flare
onset and the appearance is due to the heating through the
underneath flare reconnection. The reconnection induced the arcade
to expand outward. In the mean time, new hot loops (depicted with
yellow arrows in panels c-e) were observed to ascend continuously
(See AIA movie online for a better view), further pushing the
arcade upward (panels c-e). The rising arcade then stretched the
overlying restraining magnetic field lines (denoted with yellow
dotted lines in panels e and f). Underneath the arcade apex, the
arcade legs and the stretched field lines started to approach each
other (denoted with red arrows in panels f-h), forming a classical
X-point configuration (panel i). We believe that a second magnetic
reconnection took place at the X-point, different in the location
and time from the earlier flare reconnection that drove the
expansion of the arcade. This second reconnection led to the
formation of a hot blob.  The hot blob separated completely from
the arcade between 11:06 and 11:07 UT, following which the hot
blob faded quickly and became difficult to identify with the AIA
131 \AA \ images. Before and during this high-lying blob-forming
magnetic reconnection, the continuous rise of the arcade stretched
itself into to a cusp-like geometry, as shown in panel i with red
dotted line near the solar limb. During the reconnection, we
observed the shrinkage of the arcade below the X-point, returning
the cusp shape to a loop-shape (see AIA Movie online for the whole
process). Therefore, we suggest that in this event there exist two
reconnections taking place at different coronal heights and
playing different roles in the development of the blob formation
and eruption. One is the low-lying flare reconnection, inducing
the heating and expansion of the arcades; the other is the
high-lying reconnection, forming and releasing the hot blob.

Existing magnetohydrodynamic (MHD) simulations in spherical
coordinates (Miki\'{c} \& Linker, 1994; Antiochos et al. 1999)
have shown that, subject to the localized shearing motion in the
photosphere, the magnetic arcade in the corona expands outward in
a process that stretches the field lines and produces a current
sheet underneath between the arcade and photospheric PIL. Magnetic
reconnection occurs in the current sheet, leading to the formation
and ejection of a blob (magnetic flux rope). Such scenario shown
in these simulations is supported by our observations, at least
qualitatively. Through carefully inspecting observations of AIA
and EUVI, which provide images from different wavelengths and
perspectives, we conclude that there are no any pre-existing or
dormant MFR features before the eruption, i.e., filament, sigmoid,
blob or hot channel as mentioned in the introduction section. What
we observed is that the arcade evolved into a hot blob (flux
rope), which is consistent with the MHD models mentioned above.
The expansion of the arcade is possibly caused by the flare
magnetic reconnection and/or new rising loops from below.
Nevertheless, we want to point out that the arcade-to-flux rope
scenario observed in this particular event may not occur in other
events, in which MFR may exist before the eruption (e.g., Zhang et
al. 2012; Cheng et al. 2011). Therefore, there is no single model
that can explain the diversity of eruptive flares in real
observations at present.

\subsection{Heating Mechanism of The Arcades And Blob}

With the DEM method mentioned above (see Song et al. 2014 for
details), we further illustrate the blob formation process using
temperature maps (Figure 4), which show the details of the
temperature distribution and evolution of the arcades and the
blob. The temperature maps show that the arcade was heated from
about 5 MK to 7 MK in the first 3 minutes (panels a-c). This is
why the arcade appeared in the AIA 131 \AA \ passband (in
Figure~3) after the flare onset. The temperature continued to
increase and approached to around 10 MK (panels e-h). Following
the onset of the second magnetic reconnection at the high-lying
X-point, the hot blob was formed (see DEM Movie online). To
further elucidate the relation between the magnetic reconnection
and the heating, we investigated the location of the HXR sources
using \textit{RHESSI} data, which had effective observations
between 11:00 and 11:15 UT for our event. \textit{RHESSI}
observations are shown as black contours at 50\%, 70\%, and 90\%
of the maximum in the 12-25 keV band (panels e-h). The X-ray
source appeared only in the lower loops, and no source was
observed in the higher arcade and the blob. It has been shown for
some events that the magnetic reconnection region should be
between double X-ray sources (e.g., Liu et al. 2008; Song et al.
2014). Therefore, we suggest that the flux-rope-forming
reconnection was weak in this event and didn't have obvious
contributions to the heating. The flare region and the arcade
should be heated mainly by the flare magnetic reconnection, which
is consistent with Song et al. (2014).

\subsection{The Driver of The CME}

Was the hot blob a driver of the observed CME? For this event, the
arcade began to rise around 10:52 UT, and the hot blob started to
form at about 10:57 UT, while the appearance of the cool LF was
around 11:06 UT as shown in Figure 1(c). This time sequence of
different features supports that the blob is the driver. However,
we did not compare the velocities between the hot blob and the
cool LF, because it became difficult to track the blob after 11:06
UT, at which time the LF started to appear. To further address the
issue of the cause of the CME, we studied the kinematic evolution
of the rising arcades and the hot blob. The heights of the arcade
and later the hot blob apex (red plus symbols shown in the middle
and bottom panels of Figure 3) are tracked in the AIA images. Note
that all heights refer to the top of the arcades (or the apex of
the hot blob at later times) from the solar surface. Their
velocities are plotted in Figure 5(a) (red solid line), along with
the \textit{GOES} SXR 1-8 \AA \ profile shown by the black solid
line. We calculated these velocities from the numerical
differentiation by using the three-point Lagrangian interpolation
of the height-time data. Note that the velocity uncertainties come
mainly from the uncertainties in the height measurement. The
measurement errors are estimated to be four pixels, \textit{i.e.},
1700 km, which are propagated to estimate the velocity errors in
the standard way. Figure 5(a) shows that the velocities increased
from 40 km s$^{-1}$ to 180 km s$^{-1}$ continuously. The velocity
variation trend is tightly coinciding with the emission variation
of the associated flare. The temporal profile of the emission
intensity integrated over the flare region (depicted with the
white box in Figure 3(i)) in AIA 131 \AA \ is also plotted in this
panel. The EUV profile is very similar to that of the SXR. The two
vertical dotted lines in Figure 5 denote the time range when the
dominant hot-blob-forming magnetic reconnection took place. We
further study the relation between the reconnection energy release
and blob acceleration. In Figure 5(b) we plot the acceleration,
\textit{RHESSI} 12-25 kev HXR count rates, and derivation of SXR
versus time. The close temporal correlation between HXR flux and
the blob acceleration supports the idea that flares and CMEs are
two strongly coupled phenomena, possibly through a mutual positive
feedback process between magnetic reconnection and MFR
acceleration (e.g., Zhang et al. 2004; Temmer et al. (2008)).

It is also interesting to note that the HXR profile (and the SXR
derivative as well) has double peaks. In particular, the second
HXR peak appeared at the same time as the high-lying blob-forming
reconnection took place. This observation further supports the
two-magnetic-reconnection scenario discussed earlier,
\textit{i.e.}, the low-lying flare reconnection and the high-lying
flux-rope-forming reconnection. The flare reconnection,
responsible for producing the M1.2 class X-ray flare, was strong
and located in a lower altitude, \textit{e.g.}, near the magnetic
loop footpoints, as indicated by the HXR source. The flux-rope
forming reconnection, occurring at the X-point between the arcade
and the hot blob, might be a weaker magnetic reconnection, because
of the weaker magnetic field near the arcade top. Nevertheless,
this reconnection can also produce energetic electrons and
demonstrate its existence through the second HXR peak. The second
magnetic reconnection is essential for the blob/flux rope
formation and eruption.

\section{SUMMARY}

An M1.2 class SXR flare and the associated CME occurred at the
southwest limb on 2013 November 21. This event provided us the
first direct observation of MFR formation process from a loop
arcade during the eruption. We summarize the formation process as
follows. The coronal arcade was heated by the underlying flare
magnetic reconnection. In the same time, the arcade expanded
upward, possibly due to the same reconnection process. The arcade
was further strengthened and pushed upward by the new ascending
loops from below. The rising of the arcade further stretched the
surrounding background magnetic field. The arcade and the
stretched magnetic field then curved-in beneath the loop apex,
leading to the formation of an X-point. Then, the field lines near
the X-point continued to approach each other and a second magnetic
reconnection was induced, leading to the formation and eruption of
the hot blob ($\sim$ 10 MK), presumably a MFR, and producing a
CME. We suggest that in this event there exist two reconnections
taking place at different coronal heights and playing different
roles in the development of the MFR formation and eruption. One is
the low-lying flare reconnection, inducing the heating and
expansion of the arcade, and the other is the high-lying magnetic
reconnection, forming and releasing the MFR.

\acknowledgments We thank the referee for constructive comments
that have greatly improved this manuscript. We also thank Y. M.
Wang, R. Liu, Q. Hu, and G. L. Huang for their valuable
discussions and S. L. Ma for her help on the radial filter
technique. SDO is a mission of NASA's Living With a Star Program.
This work is supported by the 973 program 2012CB825601, NNSFC
grants 41274177, 41104113, 41274175, and 41331068. J. Zhang is
supported by NSF grant ATM-0748003, AGS-1156120 and AGS-1249270.

\clearpage

\begin{figure}
\epsscale{1.0} \plotone{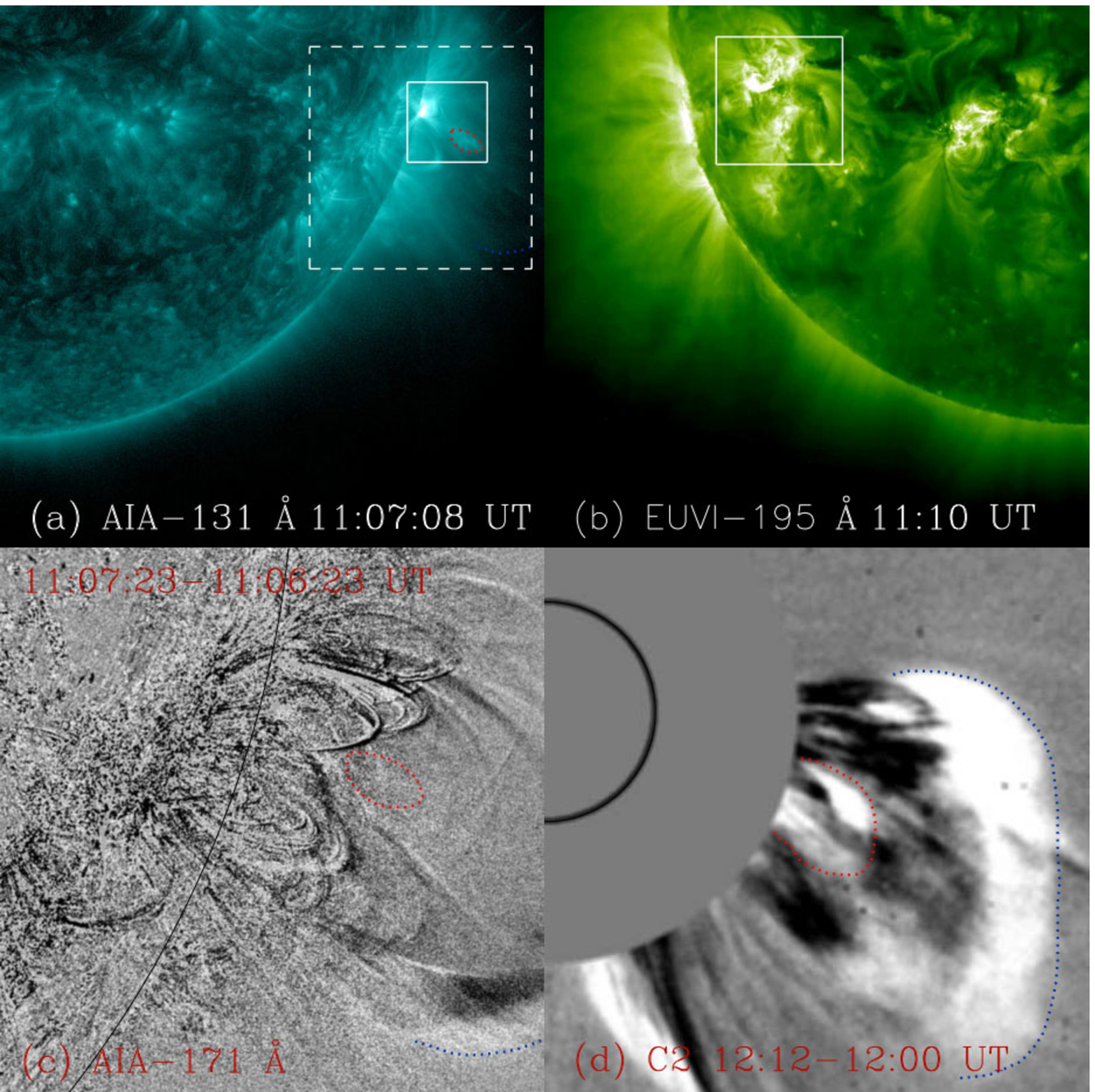} \caption{ The flare-CME event on
2013 November 21. (a) AIA 131 \AA \ image in the FOV of [0,1.3]
and [-1.3, 0] R$_\odot$ along the horizontal and vertical
directions respectively. (b) EUVI A 195 \AA \ image with FOV
[-1.3, 0] and [-1.3, 0] R$_\odot$. (c) Difference image of AIA 171
\AA. The FOV is taken to be [700, 1200] and [-600, -100] arcsec,
as depicted by the dashed square in panel (a). (d) Difference
image of LASCO/C2 in the FOV of [0, 5] and [-3.5, 1.5] R$_\odot$.
(A color version of this figure is available in the online
journal)\label{fig1}}
\end{figure}

\begin{figure}
\epsscale{0.8} \plotone{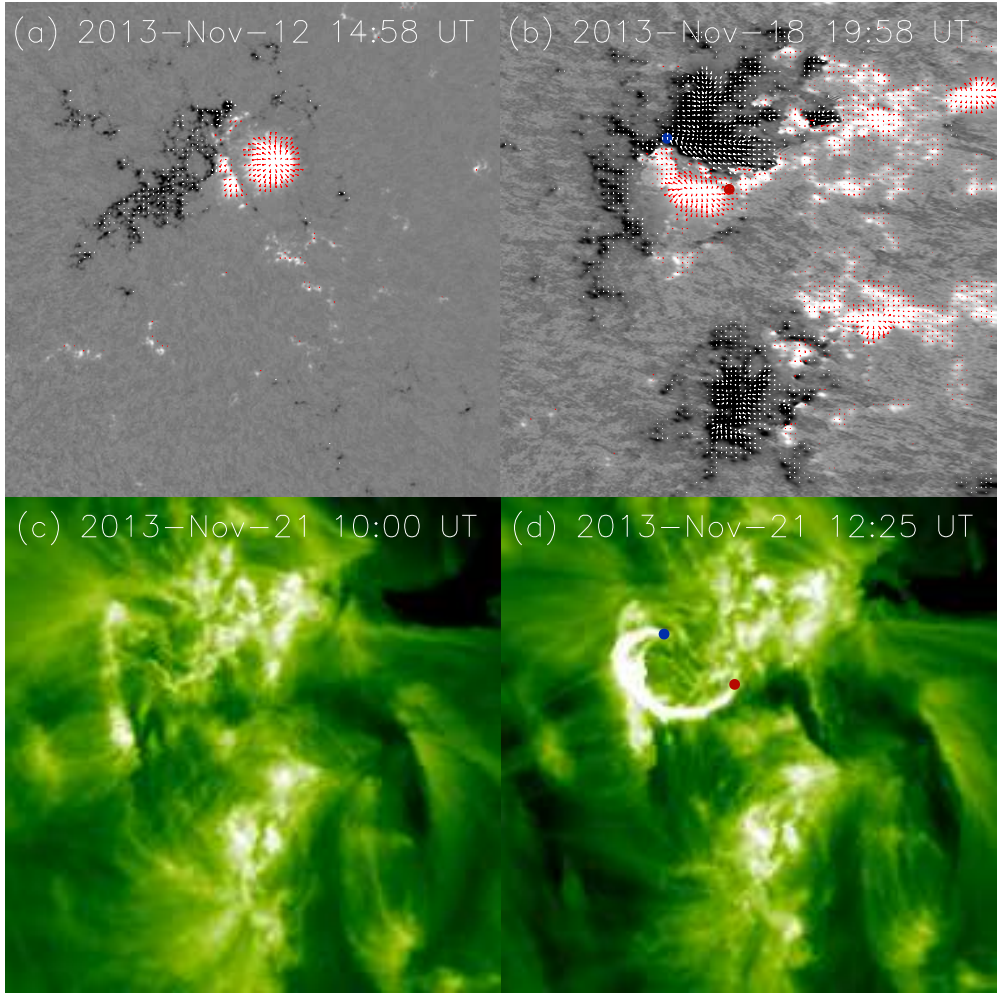} \caption{ The source region of
the flare-CME event on 2013 November 21. (a), (b) HMI vector
magnetograms. (c), (d) EUVI A 195 \AA \ images with the FOV
depicted by the solid square in Figure 1(b). (Animations (HMI.mp4
and EUVI.mp4) and a color version of this figure are available in
the online journal.)\label{fig2}}
\end{figure}

\begin{figure}
\epsscale{1.0} \plotone{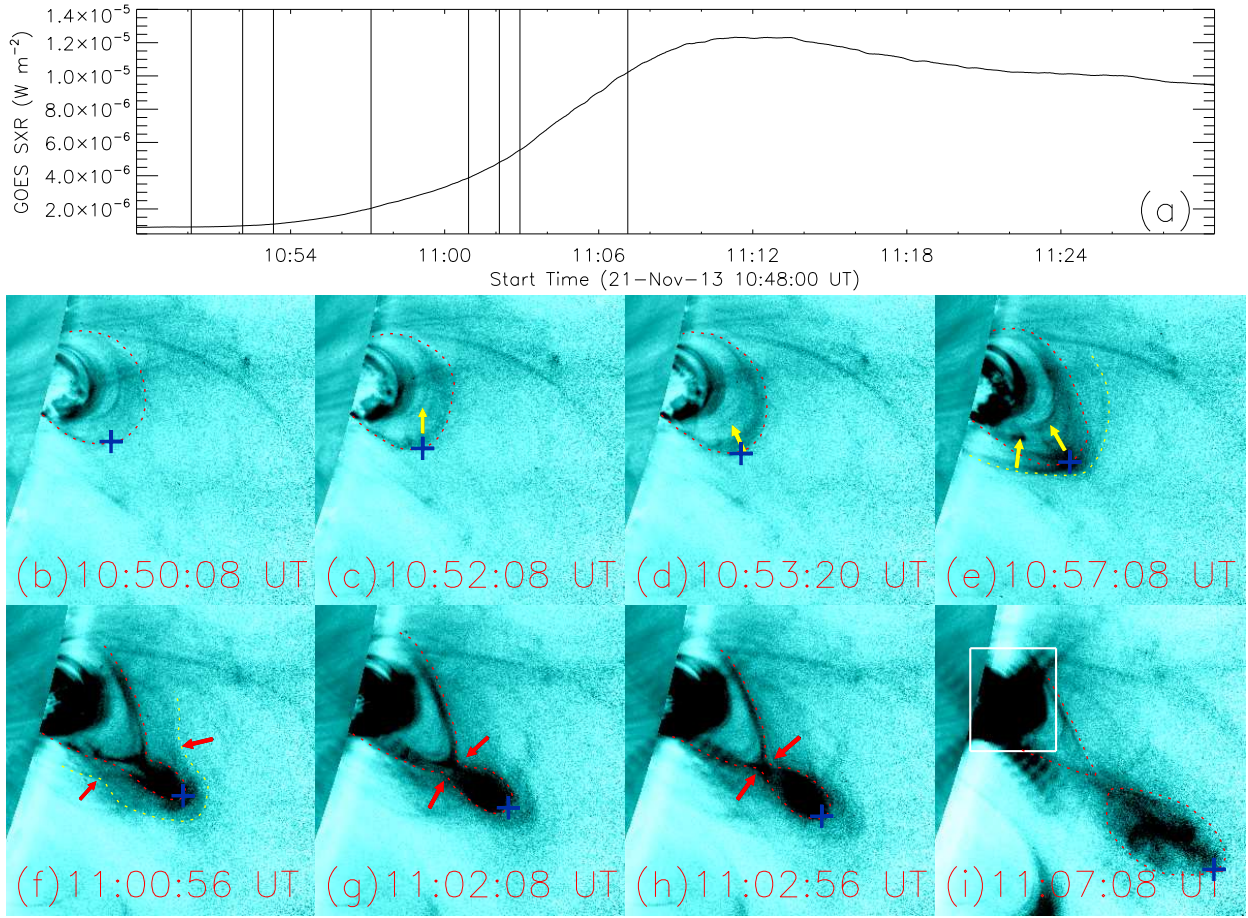} \caption{ The formation of the
hot blob. (a) The GOES SXR flux with time. Panels (b)-(i) are
snapshots of the hot blob formation process (reversed color table)
in AIA 131 \AA\ . The FOV is taken to be [920,1100] and
[-360,-180] arcsec, as depicted by the solid square in Figure
1(a). (Animation (AIA.mp4) and a color version of this figure are
available in the online journal.)\label{fig2}}
\end{figure}

\begin{figure}
\epsscale{0.8} \plotone{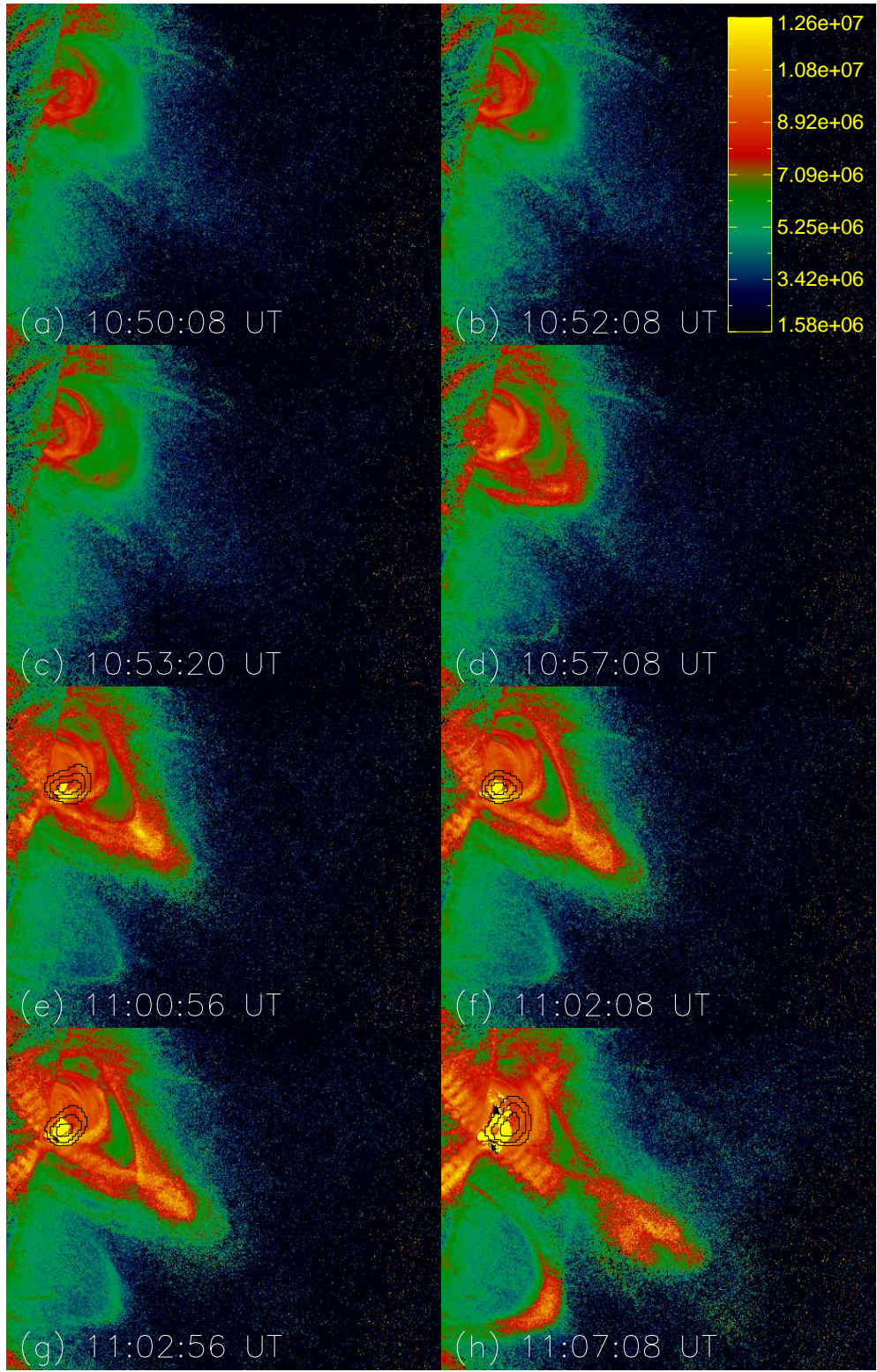} \caption{ The temperature
evolution of the eruption process of the event on 2013 November
21. The FOV is [920,1200] by [-400,-180] arcsec. (Animation
(DEM.mp4) and a color version of this figure are available in the
online journal.)\label{fig3}}
\end{figure}

\begin{figure}
\epsscale{1.0} \plotone{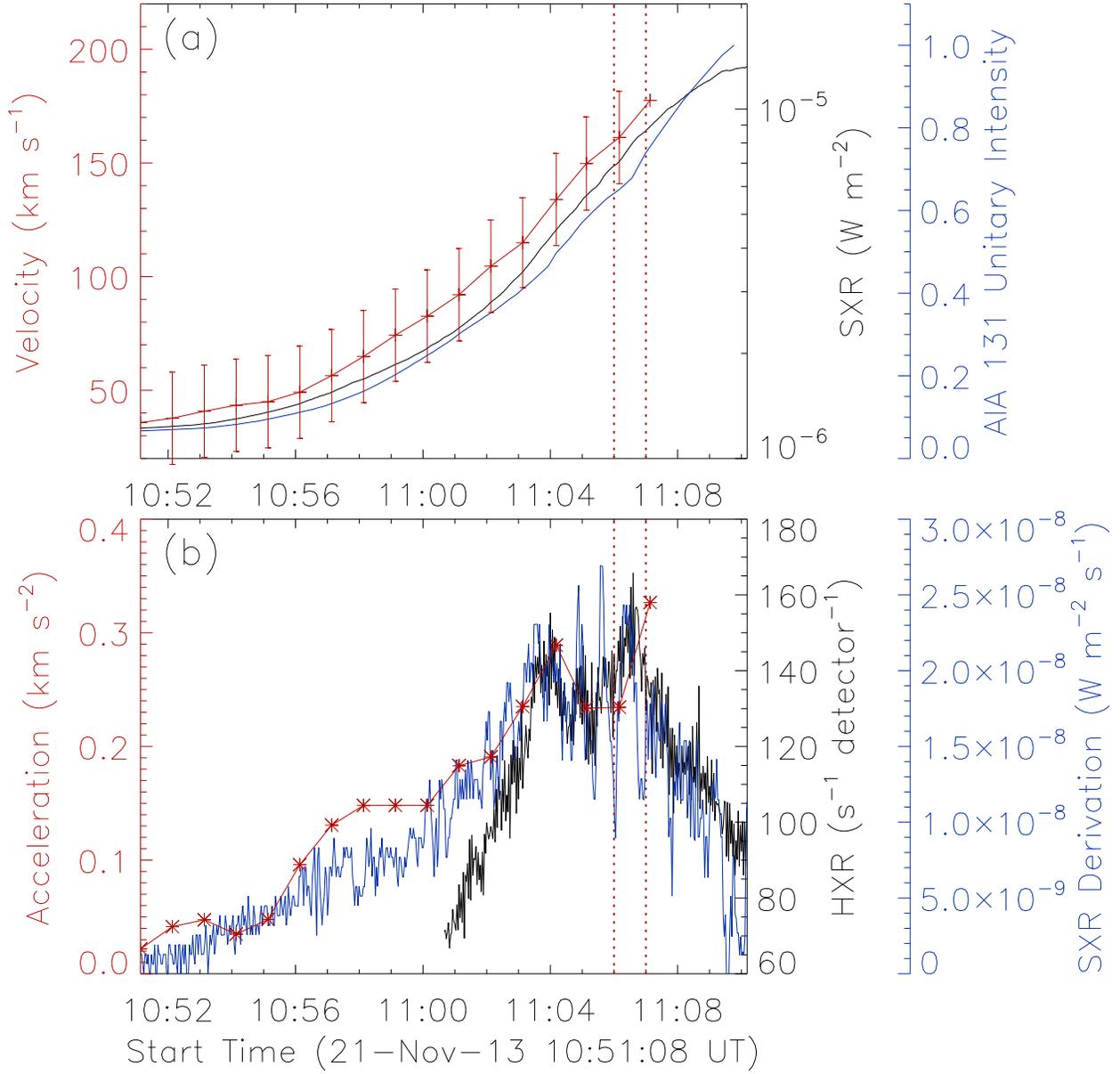} \caption{ Kinematic evolution of
the eruption. (a) The velocity-time profile of the hot blob (red),
along with the profile of \textit{GOES} SXR 1-8 \AA \ flux (black)
and AIA 131 \AA \ intensity (green). (b) The acceleration-time
profile (red), along with the intensity profiles of HXR (black)
and derivation of the SXR (green). The error bars of acceleration
are large and not shown here. (A color version of this figure is
available in the online journal.)\label{fig4}}
\end{figure}


\begin{thebibliography}{}


\bibitem[Antiochos et al. (1999)]{Antiochos 1999}
Antiochos, S. K., Devore, C. R., \& Klimchuk, J. A. 1999, \apj,
510, 485




\bibitem[Brueckner et al. (1995)]{Brueckner 1995}
Brueckner, G. E., Howard, R. A., Koomen, M. J., et al. 1995,
\solphys, 162, 357

\bibitem[Burlaga et al. (1981)]{Burlaga 1981}
Burlaga, L., Sittler, E., Mariani, F., \& Schwenn, R. 1981, \jgr,
86, 6673


\bibitem[Cheng (1996)]{Chen 1996}
Chen, J. 1996, \jgr, 101, 27499



\bibitem[Chen et al. (2006)]{Chen 2006}
Chen, Y., Li, G. Q., \& Hu, Y. Q. 2006, \apj, 649, 1093


\bibitem[Chen et al. (2007)]{Chen 2007}
Chen, Y., Hu, Y. Q., \& Sun, S. J. 2007, \apj, 665, 1421

\bibitem[Cheng et al. (2014)]{Cheng 2014}
Cheng, X., Ding, M. D., Guo, Y., et al. 2014, \apj, 780, 28

\bibitem[Cheng et al. (2013a)]{Cheng 2013a}
Cheng, X., Zhang, J., Ding, M. D., et al. 2013, \apj, 763, 43

\bibitem[Cheng et al. (2013b)]{Cheng 2013b}
Cheng, X., Zhang, J., Ding, M. D., et al. 2013, \apj, 769, L25


\bibitem[Cheng et al. (2011)]{Cheng 2011}
Cheng, X., Zhang, J., Liu, Y., \& Ding, M. D. 2011, \apj, 732, L25

\bibitem[Cheng et al. (2012)]{Cheng 2012}
Cheng, X., Zhang, J, Saar, S. H., \& Ding, M. D. 2012, \apj, 761,
62

\bibitem[Del Zanna et al. (2011)]{Del Zanna 2011}
Del Zanna, G., O'Dwyer, B., \& Mason, H. E. 2011, \aap, 535, A46



\bibitem[Forbes (2000)]{Forbes 2000}
Forbes, T. G. 2000, \jgr, 105, 23153

\bibitem[Forbes \& Isenberg (1991)]{Forbes 1991}
Forbes, T. G., \& Isenberg, P. A. 1991, \apj, 373, 294








\bibitem[Howard et al. (2008)]{Howard 2008}
Howard, R. A., Moses, J. D., Vourlidas, A., et al. 2008, \ssr,
136, 67

\bibitem[Hu et al. (2003)]{Hu 2003}
Hu, Y. Q., Li, G. Q., \& Xing, X. Y. 2003, \jgr, 108, 1072


\bibitem[Isenberg et al. (1993)]{Isenberg 1993}
Isenberg, P. A., Forbes, T. G., \& Demoulin, P. 1993, \apj, 417,
368






\bibitem[Kliem \& T\" or\" ok (2006)]{Kliem 2006}
Kliem, B., \& T\" or\" ok, T. 2006, \prl, 96, 255002

\bibitem[Klimchuk (2001)]{Klimchuk 2001}
Klimchuk, J. A. 2001, in Space Weather, ed. Song, P., Singer, H.,
\& Siscore, G. (Geophysical Monograph 125; Washington, DC: Am.
Geophys. Un.), 143




\bibitem[Lemen et al. (2012)]{Lemen 2012}
Lemen, J. R., Title, A. M., Akin, D. J., et al. 2012, \solphys,
275, 17


\bibitem[Li \& Zhang (2013)]{Li 2013}
Li, L. P., \& Zhang, J. 2013, \aap, 552, L11

\bibitem[Lin et al. (2002)]{Lin 2002}
Lin, R. P., Dennis, B. R., Hurford, G. J. et al. 2002, \solphys,
210, 3


\bibitem[Liu et al. (2008)]{Liu 2008}
Liu, W., Petrosian, V., Dennis, B. R., \& Jiang, Y. W. 2008, \apj,
676, 704



\bibitem[Ma et al. (2011)]{Ma 2011}
Ma, S. L., Raymond, J. C., Golub, L., et al. 2011, \apj, 738, 160

\bibitem[McKenzie \& Canfield (2008)]{McKenzie 2008}
McKenzie, D. E., \& Canfield, R. C. 2008, \aap, 481, L65

\bibitem[Miki\'{c} \& Linker (1994)]{Mikic 1994}
Miki\'{c}, Z., \& Linker, J. A. 1994, \apj, 430, 898

\bibitem[O'Dwyer et al. (2010)]{O'Dwyer 2010}
O'Dwyer, B., Del Zanna, G., Mason, H. E., Weber, M. A.,
\&Tripathi, D. 2010, \aap, 521, A21

\bibitem[Olmedo \& Zhang (2010)]{Olmedo 2010}
Olmedo, O., \& Zhang, J. 2010, \apj, 718, 433

\bibitem[Patsourakos et al. (2013)]{Patsourakos 2013}
Patsourakos, S., Vourlidas, A., \& Stenborg, G. 2013, \apj, 764,
125




\bibitem[Rust \& Kumar (1994)]{Rust 1994}
Rust, D. M., \& Kumar, A. 1994, \solphys, 155, 69









\bibitem[Song et al. (2014)]{Song 2014}
Song, H. Q., Zhang, J., Cheng, X., et al. 2014, \apj, 784, 48


\bibitem[Su et al. (2011)]{Su 2011}
Su, Y., Surges, V., van Ballegooijen, A., DeLuca, E., \& Golub, L.
2011, \apj, 734, 53



\bibitem[Temmer et al. (2008)]{Temmer 2008}
Temmer, M., Veronig, A. M., Vr\v snak, B., et al. 2008, \apj, 673,
L95.

\bibitem[Titov \& D\'{e}moulin (1999)]{Titov 1999}
Titov, V. S., \& D\'{e}moulin, P. 1999, \aap, 351, 707


\bibitem[T\" or\" ok et al. (2004)]{Torok 2004}
T\" or\" ok, T., Kliem, B, \& Titov, V. S. 2004, \aap, 413, L27


\bibitem[Vourlidas et al. (2012)]{Vourlidas 2012}
Vourlidas, A., Lynch, B. J., Howard, R. A., \& Li, Y. 2013,
\solphys, 284, 179

\bibitem[Wang \& Stenborg, G. (2010)]{Wang 2010}
Wang, Y. M., \& Stenborg, G. 2010, \apj, 719, L181







\bibitem[Zhang et al. (2012)]{Zhang 2012}
Zhang, J., Cheng, X., \& Ding, M. D. 2012, NatCo, 3, 747



\bibitem[Zhang et al. (2004)]{Zhang 2004}
Zhang, J., Dere, K. P., Howard, R. A., \& Vourlidas, A. 2004,
\apj, 604, 420



\end{thebibliography}
\end{document}